\documentclass[prl,twocolumn,showpacs,amsmath,amssymb,superscriptaddress,floatfix]{revtex4}

\usepackage[english]{babel}
\usepackage{graphicx}

\usepackage{times}

\begin{document}

\title{Quantum phase transitions 
in the systems of parallel quantum dots}

\author{Rok \v{Z}itko}
\affiliation{J. Stefan Institute, Ljubljana, Slovenia}

\author{Janez \surname{Bon\v ca}}
\affiliation{Faculty of Mathematics and Physics, University of
Ljubljana, Ljubljana, Slovenia}
\affiliation{J. Stefan Institute, Ljubljana, Slovenia}

\date{\today}

\begin{abstract}
We study the low-temperature transport properties of the systems of
parallel quantum dots described by the $N$-impurity Anderson model. We
calculate the quasiparticle scattering phase shifts, spectral
functions and correlations as a function of the gate voltage for $N$
up to 5.  For any $N$, the conductance at the particle-hole symmetric
point is unitary. For $N \geq 2$, a transition from ferromagnetic to
antiferromagnetic impurity spin correlations occurs at some gate
voltage.  For $N \geq 3$, there is an additional transition due to an
abrupt change in average impurity occupancy. For odd $N$, the
conductance is discontinuous through both quantum phase transitions,
while for even $N$ only the magnetic transition affects the
conductance. Similar effects should be experimentally observable in
the systems of quantum dots with ferromagnetic
conduction-band-mediated inter-dot exchange interactions.
\end{abstract}

\pacs{72.10.Fk, 72.15.Qm, 73.63.Kv}

\maketitle

\newcommand{\vc}[1]{\boldsymbol{#1}}
\newcommand{\ket}[1]{|#1\rangle}
\newcommand{\bra}[1]{\langle #1 |}

Parallel double quantum dots in semiconductor heterostructures exhibit many
interesting quantum effects at low temperatures, such as the formation of
molecular states, Aharonov-Bohm oscillations, phase lapses and the Kondo
effect \cite{holleitner2004, hatano2005, sigrist2004, sigrist2006, chen2004,
craig2004, sasaki2006}. They are also predicted to exhibit quantum phase
transitions of different kinds \cite{vzporedne, dasilva2006}.  Inter-dot
exchange interactions, both ferromagnetic (FM) and antiferromagnetic (AFM),
play a central role in such systems \cite{lopez2002, lopez2005, utsumi2004,
simon2005, simon2005prb, tamura2005, vavilov2005, chiappe2004}. Furthermore,
the phase coherent electron transport leads to various possible interference
effects and Fano anti-resonances \cite{lu2005, ladron2003, ladron2006,
karrasch2006}. Recently, few-electron triple quantum dot structures have
been fabricated \cite{gaudreau2006, korkusinski2007prb} and even more
complex multi-dot nanostructures can in principle also be assembled. It is
thus appropriate to endeavor the studies of new phenomena which occur in
multiple-dot systems.

\begin{figure}[htbp]
\centering
\includegraphics[width=8cm]{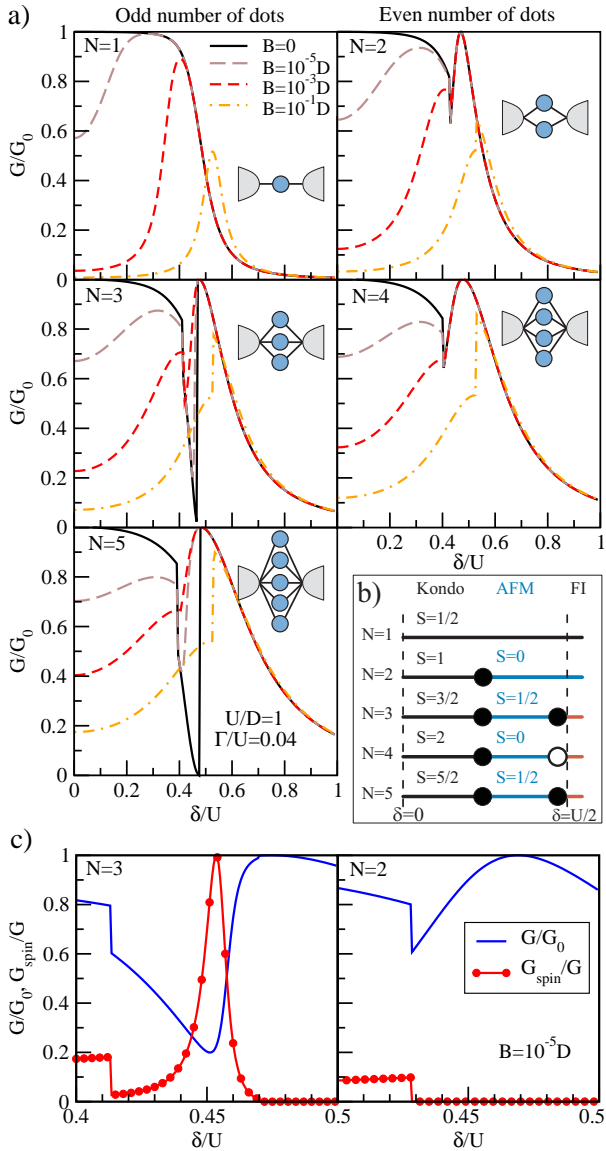}
\caption{(color online) a) Zero-temperature conductance through
systems of $N$ parallel quantum dots as a function of the gate voltage
for a range of magnetic fields. The in-plane field only leads to
Zeeman splitting; no magnetic flux pierces the rings formed by pairs
of dots.  $\Lambda=4$; NRG iterations were performed until the zero
temperature limit was reached.
$T_K=3.3\ 10^{-6}D$ (for $N=2$). The magnetic field is measured in
units of $g\mu_B$. Only $\delta>0$ is shown due to the symmetry of the
problem.
b) Zero-temperature phase diagram delimiting the different regimes as a
function of the gate voltage. Filled circles ($\bullet$) correspond to phase
transitions visible in conductance, while the empty circle ($\circ$)
denotes the phase transition with no associated conductance discontinuity.
c) Conductance and spin conductance
 in small magnetic field in the
transition region.
}
\label{figsfield}
\end{figure}

The simplest $N$-impurity Anderson model for several dots embedded in
parallel between two conduction leads in a left-right symmetric way
(see the insets in Fig.~\ref{figsfield}a) is defined by
$H=H_b+\sum_{i=1}^{N} H_i$, where $H_\mathrm{b}$ describes a band with
a constant density of states $\rho=1/2D$ ($2D$ is the bandwidth) and
\begin{equation}
H_i = \delta n_i + \frac{U}{2} (n_i-1)^2 + 
V \sum_{k\sigma} (c^\dag_{k\sigma} d_{i\sigma} + \text{H.c.}).
\end{equation}
Parameter $\delta=\epsilon+U/2$ is related to the gate voltage, $U$ is the
electron-electron repulsion and we assume that all impurities hybridize with
the same left-right symmetric combinations of states from both leads with a
constant hybridization function $\Gamma=\pi \rho V^2$; this corresponds to
taking the limit of small inter-dot separation \cite{vzporedne}. The
inter-dot tunneling coupling and capacitive coupling (inter-dot charge
repulsion) are assumed small and all dots are equivalent: the system thus
has $S_N$ symmetric group symmetry of all possible permutations of dot
labels $i$. At the particle-hole (p-h) symmetric point, $\delta=0$, and for
$U/\pi \Gamma \gg 1$, the conduction-band-mediated inter-dot exchange
interaction induces FM alignment of the impurity spins and the system
undergoes the spin-$N/2$ Kondo effect ending up in an underscreened
strong-coupling (SC) Fermi liquid fixed point with residual spin $N/2-1/2$
\cite{vzporedne, jayaprakash1981, hofstetter2002, tamura2005, utsumi2004,
posazhenikova2005, lopez2005, pustilnik2006}. For very large $\delta/U$, the
impurities are unoccupied and the system is in the frozen-impurity (FI)
fixed-point with no residual spin. In the single-impurity ($N=1$) case, the
SC and FI fixed points lie on the same line of fixed points and they differ
only in the strength of the potential scattering \cite{krishna1980b}. For $N
\geq 2$, however, the SC and FI lines of fixed points are qualitatively
different (each corresponding to a different residual spin) and must be
separated by at least one quantum phase transition (QPT) \cite{vojta2006}.

A reliable technique to study the low-temperature properties of
coupled quantum dot systems is the numerical renormalization group
(NRG) \cite{wilson1975, bulla2007}. %
In this Letter, we show that for any $N \geq 2$, the FM alignment
collapses at some critical value $\delta_{c1}$ and that for $\delta >
\delta_{c1}$ the inter-impurity spin-spin correlations are AFM. For $N
\geq 3$, there is precisely one additional QPT at slightly higher
$\delta_{c2}$ related to an abrupt change in the average impurity
occupancy. The essential novelty of our results is that certain of
these phase transitions can be easily detected in zero-bias
conductance measurements.

{\it Conductance.} -- The on-site energy $\epsilon$ can be regulated using
the gate voltages to tune the charge state (occupancy) on the dots.
Gate-voltage dependent conductance is shown in Fig.~\ref{figsfield}a for
$N=1,\ldots,5$ for a range of magnetic field strengths. The conductance is
calculated as $G=G_0/2 \sum_{\sigma=\uparrow,\downarrow} \sin^2
{\delta_{\mathrm{q.p.}}^\sigma}$ where $G_0=2e^2/h$ is the conductance
quantum \cite{pustilnik2001} and the quasiparticle scattering phase shifts
$\delta_{\mathrm{q.p.}}^{\sigma}$ are extracted from the NRG excitation
spectra.

At $\delta=0$, the systems are fully conducting at zero field and
there is a wide plateau of high conductance associated with the
spin-$N/2$ Kondo effect \cite{vzporedne}. While the $N=1$ system
smoothly crosses over from the Kondo regime to the non-conducting FI
regime, in the multi-impurity case we observe sharp discontinuities:
one discontinuity for $N$ even and two discontinuities for $N$
odd. The conductance culminates in a unitary peak slightly below
$\epsilon=0$ (i.e. below $\delta/U=1/2$) for all $N \geq 2$.  The
origin of this peak is simply potential scattering. The magnetic field
$B$ has a strong effect on the Kondo plateau: the conductance is
significantly reduced as soon as $B$ is of the order of the Kondo
temperature $T_K$. The potential scattering peak is only affected by
extremely high fields of the order of $U$.

\begin{figure}[htbp!]
\includegraphics[clip,width=8cm]{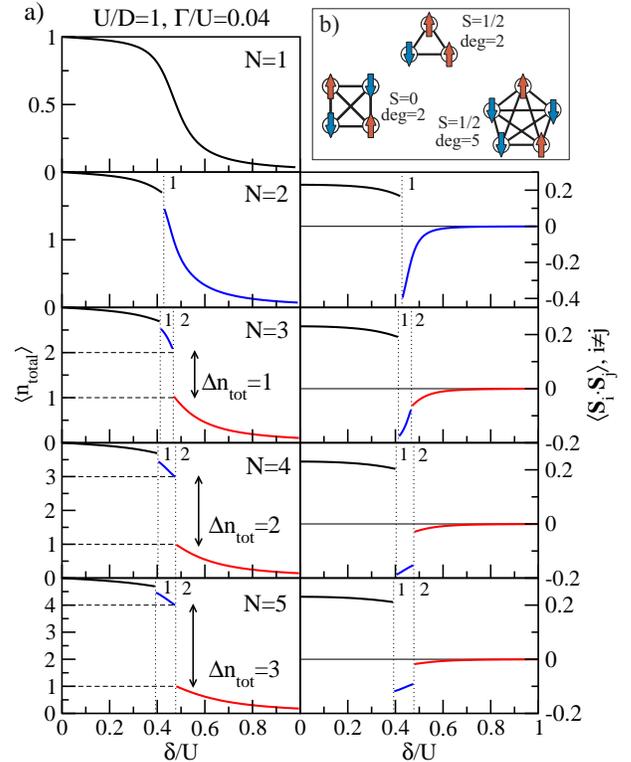}
\caption{(color online) a) Total occupancy (charge) and spin-spin correlation between
pairs of spins 
as a function of the gate voltage. b) Schematic representations of the
spin configurations in the intermediate AFM ordered phases, their total
spins and degeneracies for $N \geq 3$.}
\label{fig_e}
\end{figure}

{\it Quantum phase transitions.} -- 
Conductance discontinuities find their counterparts in the jumps of the
total impurity occupancy and spin-spin correlation $\langle \vc{S}_i \cdot
\vc{S}_j \rangle$ (Fig.~\ref{fig_e}a); a new feature, however, is the
existence of two points of discontinuity for $N=4$ while the conductance
only exhibits one. In the Kondo regime for $\delta < \delta_{c1}$, the
systems are nearly half-filled and spins are aligned \cite{vzporedne}. As we
cross $\delta_{c1}$, the occupancy abruptly decreases and the spin
correlations turn from FM to AFM. For $N \geq 3$, a second discontinuity
occurs at somewhat higher $\delta_{c2}$; its characteristic property is the
occupancy jump by almost exactly $N-2$, from $N-1$ to $1$. According to the
Friedel sum rule, a change in the occupancy by $n$ is mirrored in a change
of the scattering phase shift by $\Delta \delta_{\mathrm{q.p.}} = n \pi/2$.
This explains the conductance jump from $G=0$ to $G=G_0$ in the case of odd
$N \geq 3$ and the absence of the second conductance discontinuity for even
$N \geq 4$. It is remarkable that the second QPT occurs precisely at the
point where the conductance is extremal.

\begin{figure}[htbp!]
\includegraphics[width=8cm,clip]{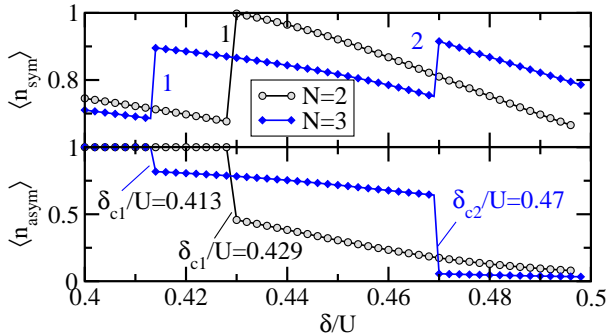}
\caption{(color online) Expectation values in the symmetric and asymmetric basis for 
$N=2$ and $3$ systems. The asymmetric combination of states considered
is $d^\dag_{\mathrm{asym}} = 1/\sqrt{2} (d^\dag_1-d^\dag_2)$.
}
\label{fig_n}
\end{figure}

The discontinuities originate from an interplay of the RKKY
interactions \cite{vzporedne}, the presence of the bound states in the
continuum \cite{ladron2003, ladron2006, lopez2005}, and the occupancy
switching \cite{silvestrov2000, sindel2005, apel2005}. We note that
only the symmetric state described by the operator
$d^\dag_{\mathrm{sym}} = 1/\sqrt{N} \sum_i d_i^\dag$ hybridizes
directly with the conduction band, while all asymmetric states are
decoupled.  A close-up on the points of discontinuity for $N=2,3$ is
shown in Fig.~\ref{fig_n}. As we move away from $\delta=0$, only the
symmetric state is depopulated until we reach $\delta_{c1}$. At this
point, the asymmetric levels are depopulated and the symmetric level
re-populated (up to exactly 1 for $N=2$); such occupancy switching
stems from the competition between the $\Gamma$-dependent coupling of
dots to the conduction band and the charging energy $U$
\cite{silvestrov2000}.  Between $\delta_{c1}$ and $\delta_{c2}$, the
occupancy of all states decreases until at $\delta_{c2}$ another
charge oscillation occurs in which the occupancy of the asymmetric
states plummets. Both transitions can be classified as first-order
boundary quantum phase transitions (level-crossing) \cite{apel2005};
the schematic phase diagram in shown in Fig.~\ref{figsfield}b.

For $\delta_{c1} \lesssim \delta$ ($N=2$) and for $\delta_{c1} < \delta <
\delta_{c2}$ ($N \geq 3$), each spin interacts with every other spin
antiferromagnetically with an equal strength. The ground state of such
effective Heisenberg Hamiltonian consists of one singlet for $N=2$, two
degenerate doublets for $N=3$, two degenerate singlets for $N=4$, etc.
\cite{vandersijs1993}, see Fig.~\ref{fig_e}b. Results for magnetic
susceptibility and entropy demonstrate that in odd-$N$ systems the spin
degree of freedom is screened by the spin-$1/2$ Kondo effect and that the
residual entropy is the logarithm of the additional degeneracy of the ground
state spin multiplets. This implies that in this phase (for $N \geq 3$)
there are several equally probable ways for the electron spin ordering.
Consequently, this ``magnetic-frustration'' phase is sensitive to the
breaking of the $S_N$ symmetry between the impurities which lifts the
degeneracy. For odd $N$ and for weak magnetic field, the transmitted current
in this regime becomes fully spin polarized at some gate voltage (it should
be noted that in the presence of magnetic field the phase transition at
$\delta_{c2}$ is replaced by a cross-over \cite{pustilnik2006}), see the
plot of $G_\mathrm{spin}/G$ where
$G_\mathrm{spin}=G_0/2(\sin^2\delta^\uparrow_\mathrm{q.p.}-\sin^2
\delta^\downarrow_\mathrm{q.p.})$ is spin current and $G$ is total (charge)
current, Fig.~\ref{figsfield}c. Thus the $N=3$ device might function as a
spin filter \cite{pustilnik2006, torio2004filter, aligia2004filter}.

For decreasing interaction strength $U$, the discontinuous features in
the conduction plots narrow down and disappear for $U=0$; at this
point the conductance curve consists of a single Lorentzian peak of
width $N \Gamma$ centered at $\delta=0$. This demonstrates that the
conductance discontinuities are adiabatically connected with the ghost
Fano resonances (bound states in the continuum) found in the systems
of noninteracting parallel dots \cite{ladron2003, ladron2006}.

{\it Robustness.} -- Finite inter-dot charge repulsion $U_{12}$ and
inter-dot hopping $t$ have little effect as long as $U_{12} \ll U$, $t
\ll U-U_{12}$ and $4t^2/(U-U_{12}) \ll J_\mathrm{eff}$, where
$J_\mathrm{eff}$ is the effective conduction-band mediated
inter-impurity exchange interaction \cite{vzporedne}. The first two
constraints are easily met in experiments \cite{hatano2004} and
together they imply the third unless $J_\mathrm{eff}$ is very small.

The $S_N$ symmetry is broken if the strengths of the hybridization
$\Gamma_i$ of each impurity to the conduction band are made unequal,
or if different gate voltages $\epsilon_i$ are applied to the dots.
This leads to smearing of the discontinuities in the occupancy and
correlation functions; nevertheless, the discontinuities in the
conductance curves persist (i.e. level-crossings still
occur). Curiously, the first discontinuity in the conductance no
longer coincides with the sign change of $\langle \vc{S}_i \cdot
\vc{S}_j \rangle$.

For a generic problem of parallel dots, in particular if left-right (L-R)
and $S_N$ symmetries are weakly broken, the antisymmetric combinations of
the conduction band electrons become relevant and we need to consider the
full $N$-impurity two-channel model. In this case, the spin-$N/2$ Kondo
effect is followed by another stage of the Kondo screening to $S=N/2-1$ at
significantly lower Kondo temperature $T_K^{(2)}$. This leads to a phase
shift of $\pi/2$ in the odd scattering channel and the zero-temperature
conductance in the ``Kondo regime'' becomes very small \cite{pustilnik2001}
(see Refs.~\cite{tamura2005, karrasch2006, pustilnik2006} for the $N=2$
case). Nevertheless, if L-R breaking is weak, the conductance curves shown
in Fig.~\ref{figsfield} are a good approximation for the finite-temperature
conductance in the experimentally relevant $T_K \gg T \gg T_K^{(2)}$ range.

\begin{figure}[htbp!]
\includegraphics[width=8cm,clip]{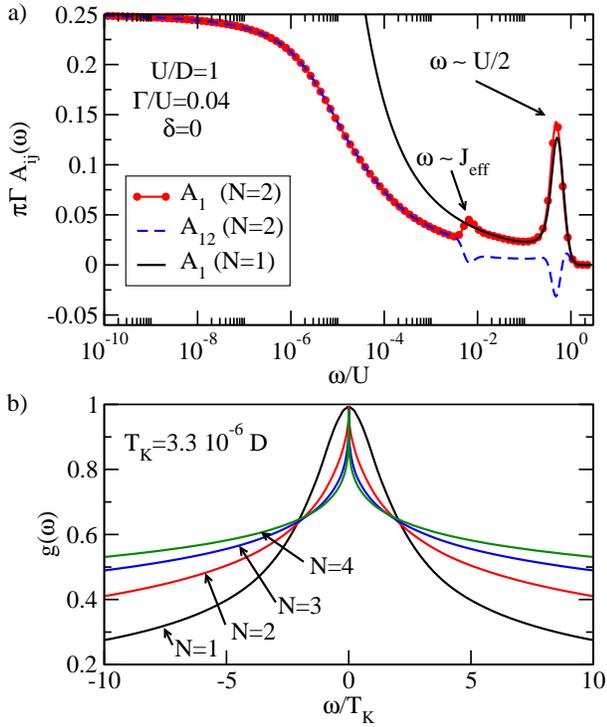}
\caption{(color online)
a) Normalized spectral functions $\pi \Gamma A_{ij}(\omega)$ and
b) function $g(\omega)=\pi \Gamma \sum_{ij} A_{ij}(\omega)$.
}
\label{fig_i}
\end{figure}

{\it Spectral functions.} -- In Fig.~\ref{fig_i}a, we plot spectral
functions $A_{ij}(\omega)=-1/(2\pi)\mathrm{Im} (G^r_{ij}+G^r_{ji})$
for $N=1$ and $2$ at the p-h symmetric point \cite{spincharge}. The
diagonal spectral functions $A_i=A_{ii}$ represent the on-site density
of states, while the out-of-diagonal spectral functions $A_{ij}$ with
$i\neq j$ are related to the processes where one electron is injected
at one site and later extracted at a different site. The peak in $A_i$
at $U/2$ is the familiar charge excitation peak that is observed for
all $N$; for $N \geq 2$ a small negative peak appears in the
out-of-diagonal spectral densities $A_{ij}$. Additional features for
$J_\mathrm{eff} \lesssim \omega \lesssim U$ are related to the
magnetic alignment \cite{flnfl3, vavilov2005}: the diagonal spectral
function exhibits a broad hump which peaks at $\sim J_\mathrm{eff}$,
while the out-of-diagonal spectral function exhibit a slight
depression. For $\omega < J_\mathrm{eff}$ when the spins align, all
$A_{ij}$ curves merge into a Kondo resonance.

In Fig.~\ref{fig_i}b, we plot the symmetrized and normalized spectral
density function $g(\omega)=\pi \Gamma \sum_{ij} A_{ij}$, which
determines the conductance through the dots as $G=G_0 \int (-\partial
f/\partial \omega) g \mathrm{d}\omega$, where $f$ is the Fermi-Dirac
distribution function \cite{meir1992, spincharge}. In a simple
approximation, $g(\omega)$ at $\omega=T$ indicates the conductance
through the system at temperature $T$.
Only for $N=1$ is the approach to the unitary conductance at $T=0$
rapid (quadratic), as expected for regular Fermi liquid systems.  For
$N \geq 2$, the Kondo resonance is cusp-like and the approach to the
unitary conductance is very slow (logarithmic)
\cite{posazhenikova2005, koller2005}. This is characteristic for
underscreened Kondo systems which behave as singular Fermi liquids
\cite{mehta2005, koller2005}.

{\it Conclusion.} -- For $N \geq 3$, the $N$-impurity Anderson model
undergoes two phase transitions. 
The first transition separates the spin alignment and the associated
spin-$N/2$ Kondo screening from the spin anti-alignment with magnetic
frustration and (for odd $N$) Kondo screening of the spin-$1/2$
moment. The second transition reflects the instability of the phases
with the occupancy in the interval $1 < \langle n_\mathrm{tot} \rangle
< N-1$. Furthermore, for odd $N$ the system abruptly switches from
being fully conducting to zero conductance; this would facilitate the
experimental observation of similar effects in quantum dot systems and
might even have applications as a switch or a transistor with very
high on-off ratio. In addition, the odd $N$ system appears as a
possible realization of gate voltage switchable spin filter device.

The authors acknowledge the financial support of the SRA under Grant
No. P1-0044.

\bibliography{vzporedne.bib}

\end{document}